\documentclass[twocolumn,showpacs,preprintnumbers,amsmath,amssymb]{revtex4}
\input{epsf}
\input epsf.sty
\input psfig.sty

\begin{document}

\title{Solitons in Trapped Bose-Einstein condensates in
one-dimensional optical lattices}

\author{F. H\'ebert$^1$, G.G. Batrouni$^1$, R. T. Scalettar$^2$}
\affiliation{$^1$Institut Non-Lin\'eaire de Nice, Universit\'e de
Nice--Sophia Antipolis, 1361 route des Lucioles, 06560 Valbonne,
France} \affiliation{$^2$Physics Department, University of California,
Davis CA 95616, USA}

\begin{abstract}
We use Quantum Monte Carlo simulations to show the presence and study
the properties of solitons in the one dimensional soft-core bosonic
Hubbard model with near neighbor interaction in traps. We show that
when the half-filled Charge Density Wave (CDW) phase is doped,
solitons are produced and quasi long range order established. We
discuss the implications of these results for
the presence and robustness of this solitonic phase in
Bose-Einstein Condensates (BEC) on one dimensional optical lattices in
traps and study the associated excitation spectrum.  
The density profile exhibits
the coexistence of Mott insulator, CDW, and superfluid regions.
\end{abstract}

\pacs{03.75Hh,03.75.Lm,05.30.Jp}

\maketitle

Localization phenomena in reduced dimensionality take place in a wide
range of systems. Shape preserving excitations like lattice solitons,
which are intrinsically localized modes, are caused by an interplay of
the discreteness of the lattice and non-linearity of the underlying
dynamics. These objects have been the focus of intense experimental
activity and have been observed in optical waveguide
arrays~\cite{fleischer03}, in the quasi one dimensional
antiferromagnetic material $({\rm C}_2{\rm H}_5{\rm NH}_3)_2{\rm
CuCl}_4$~\cite{sievers98}, and in quasi-one-dimensional Bose-Einstein
condensates (BEC)~\cite{burger99}. In this last example, a $\Delta
\phi=\pi$ phase flip is imposed on a BEC with repulsive interactions
(positive scattering length), exciting a dark soliton.  Topological
excitations in lattice models of polyacetylene also have a long
history~\cite{su79}.  There, localized regions exist in which a
transition occurs between two possible configurations of long and
short bonds.  For reviews, see \cite{fisher89,bronski01}.

Recent effort has focussed on the existence of such localized modes in
BEC.  Dark solitons were exhibited in numerical solutions to the
Gross-Pitaevskii equation~\cite{burger99,trombettoni01} while bright
solitons were shown to exist in BEC with attractive interactions
(negative scattering length) by solving numerically the non-linear
Schr\"odinger equation~\cite{ruprecht95}. In addition, bright solitons
are known to exist for repulsive interactions when the effective mass
is negative~\cite{oberthaler02} and were shown to exist experimentally
for positive scattering length atomic condensates on trapped optical
lattices~\cite{oberthaler04}.  
Since condensate interactions and lattice parameters can be precisely tuned,
optical lattices offer the possibility to
explore systematically exotic soliton phases.
Variational calculations~\cite{ahufinger04,odell03} have begun to explore the
dynamics and excitations of such systems.

In this paper we use Quantum Monte Carlo (QMC) simulations to
determine the effect of near neighbor (nn) repulsive interactions on
the ground state phase diagram of BEC on 1d optical lattices, both
with and without traps.  Our model is the bosonic Hubbard tight
binding model,
\begin{eqnarray}
H&=&-t\sum_{i} \left(a^{\dagger}_i a_{i+1} + a^{\dagger}_{i+1}
a_i\right) + V_T\sum_i x^2_i \, n_i\\ & & + U\sum_i n_i(n_i-1) +
V_1\sum_{i} n_i n_{i+1}. \nonumber
\label{hubham}
\end{eqnarray}
The hopping parameter, $t$, sets the energy scale, $n_i=a^\dagger_i
a_i$ is the number operator,
$[a_i,a^\dagger_j]=\delta_{ij}$ are bosonic creation and destruction
operators. $V_T$ sets the confining trap curvature, while the contact
and near-neighbor interactions are given by $U$ and $V_1$. We use the
World Line algorithm and the Stochastic Series Expansion (SSE) method,
which work in the canonical and grand canonical ensembles,
respectively, and simulate both the soft core ($U$ finite) and hard
core cases.

The phase diagram with $V_T=0$ is known at half
filling~\cite{niyaz94}. For $V_1< 2t$ the ground state is superfluid
while for large $U$ and $V_1>2t$ off diagonal long range order is
replaced by an incompressible, insulating charge density wave (CDW)
phase where sites alternate between high and low occupation. Away from
half and integer filling the system is always superfluid. In the
hardcore limit, this model can be mapped onto a spinless 1d fermionic
model with near-neighbor repulsion and also onto the quantum
spin-$1/2$ XXZ model.  The extended 1d fermion Hubbard model (with
spin) and the classical spin chain model are both known to have
soliton excitations~\cite{sievers98,sandvik02}.  The main focus of
this paper is to demonstrate that solitons continue to exist despite
the possibility of multiple occupancy and the presence of a trap.

Static and dynamic quantities like the density and compressibility
have already been shown to exhibit unusual features due to the trap,
requiring local generalizations of these global
quantities.\cite{batrouni02,wessel04} We also show that even more
complex spatial structures can exist when the near-neighbor repulsion
is nonzero.

To address the question of solitonic excitations, we measure the
structure factor at equal imaginary time,
\begin{equation}
S(k) = \frac{1}{L^2} \sum_{x,x^\prime} {\rm e}^{ik(x-x^\prime)}\langle
n(x,\tau)n(x^\prime,\tau) \rangle,
\label{sk}
\end{equation}
where $L$ is the number of lattice sites and $0\leq \tau \leq \beta$.
Our simulations are done at $\beta=10$ which is large enough to study
the ground state.  To make contact with the excitation spectrum, we
use the $f$-sum rule
\begin{equation}
\int_{-\infty}^{+\infty} {\rm d}\omega\,\, \omega \, S(k,\omega) = N_b E_k,
\label{fsum}
\end{equation}
where $S(k,\omega)$ is the dynamic structure factor ($N_b S(k)= \int
{\rm d}\omega S(k,\omega)$) and
\begin{equation}
 E_k = \frac{-t}{L}\left ( {\rm cos}\,k-1\right ) 
\langle 0| \sum_{i=1}^L
 \left ( a^{\dagger}_i a_{i+1} + a^{\dagger}_{i+1} a_i\right)|0 \rangle,
\end{equation}
The dispersion relation is given by the Feynman result,
\begin{equation}
\Omega(k) = \frac{E_k}{S(k)}.
\label{disp}
\end{equation}
The dispersion curves shown below are obtained with
Eq.~\ref{disp}. However, we have verified that we obtain the same
results by measuring the imaginary-time-separated density-density
correlation function, performing the Laplace transform using the
maximum entropy algorithm to obtain $S(k,\omega)$, and applying
Eq.~\ref{fsum} directly.

\begin{figure}
\psfig{file=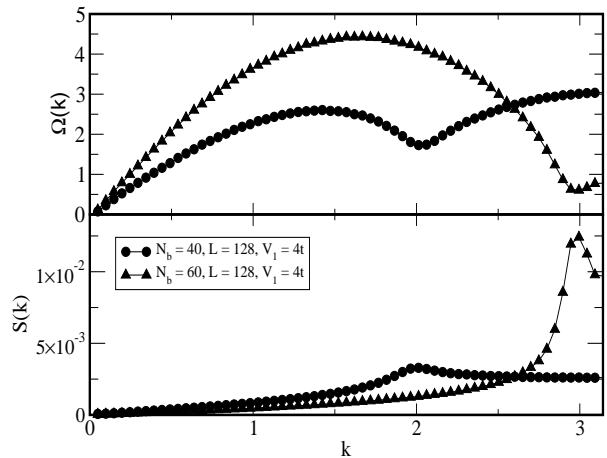,height=3in,width=3.5in,angle=-90}
\vskip-0.3in
\caption{The dispersion relation, $\Omega(k)$ and the static structure
factor, $S(k)$, vs $k$.  Peaks in $S(k)$ at incommensurate wavevectors
result in the soliton minima in $\Omega(k)$.}
\label{hardomega}
\end{figure}

We first address the uniform, $V_T=0$, system in the hardcore limit
using the SSE algorithm.  Figure~\ref{hardomega} shows $S(k)$ and
$\Omega(k)$ for $L=128$ sites and two values of $N_b$ off half
filling. Note that for small $k$, $\Omega(k)$ for both fillings
behaves linearly, indicating phonon excitations and superfluid
stability under the Landau criterion.  The velocity of sound can be
extracted from this small $k$ behavior.  Furthermore, the peak at
$S(k^*)$, observed at half filling for $k^*=\pi$, and which indicates
the CDW order, does not vanish when the system is doped. Instead, its
height decreases and $k^*$ shifts to lower values. Corresponding to
this quasi-long range order peak in $S(k)$ is a dip in $\Omega(k=k^*)$
giving the soliton excitation energy. This is similar to the roton
minimum in two and three dimensions. As the system is doped further,
the soliton minimum, and thus quasi-long range order, will disappear.
The solitons are also evident in the real space boson density (not
shown) in the form of well-localized regions of cross over between the
two possible sublattices holding high and low density sites.  The
presence of soliton excitations in the hardcore system is in agreement
with what is known for fermionic and classical spin chains.

\begin{figure}
\psfig{file=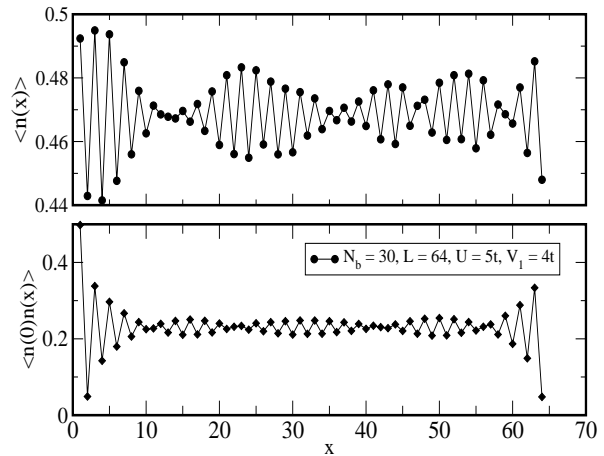,height=3in,width=3.5in,angle=-90}
\vskip-0.3in
\caption{The density profile and correlation function (averaged over
 $10^6$ QMC sweeps) for the doped soft core system.}
\label{soft30profcorr}
\end{figure}

\begin{figure}
\psfig{file=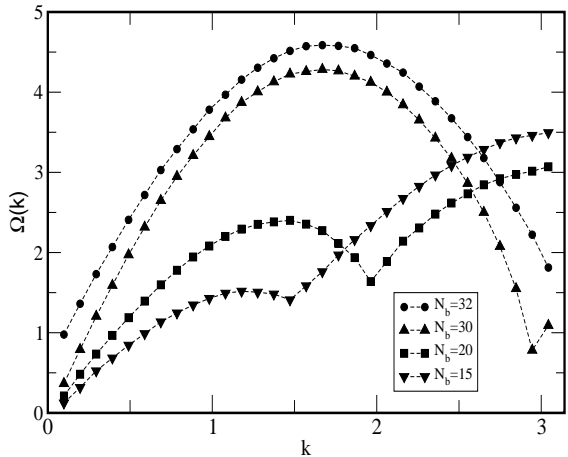,height=3in,width=3.5in,angle=-90}
\vskip-0.3in
\caption{The dispersion relation, $\Omega(k)$, vs $k$ for several
  fillings for the soft core model, $U=5,\,V_1=4$ at $\beta=10$ and $L=64$.
Solitons survive despite multiple occupancy.}
\label{softomega}
\end{figure}

The crucial question is whether this behaviour persists for the soft
core case which is relevant to atomic condensates on optical
lattices. In the soft core case, the contact interaction, $U$, must be
large enough to suppress multiple occupancy in order to stabilize the
CDW phase at half filling when $V_1$ is large enough.\cite{niyaz94}
Such large values of $U$ have been achieved experimentally on optical
lattices and lead to the Mott phase at full filling. In what follows
we fix $U=5t$.  At $V_1=4t$, the ground state density profile and
correlation function at half filling exhibit a strong CDW pattern.
The density-density correlations oscillate with nearly maximal
amplitude, indicating quantum fluctuations are small, and show little
decay with increasing separation.  Doping this system by removing two
bosons (Fig.~\ref{soft30profcorr}) yields pronounced, long-lived,
soliton excitations.  In real space, as seen in Fig.~2, these appear
as local regions of density alternation modulated by an overall
`beating' pattern.  The beat wavelength (soliton size) is given by
$\Delta x= 2 \pi / (\pi - k_*)$, where $k_*$ is the position of the
soliton energy minimum in $\Omega(k)$.  In Fig.~\ref{softomega} we
show the dispersion $\Omega(k)$ for several fillings. $S(k)$ behaves
like the hardcore case with a peak at $k=k^*$ corresponding to the
soliton minimum which moves towards lower $k^*$ as doping is
increased.  As in the hard-core case, for $N_b<32$, where the system
is superfluid, $\Omega(k)\propto k$ for small $k$ showing the
stability of the superfluid and the absence of a gap. However, for
$N_b=32$, the system is a gapped CDW insulator.  This is seen clearly
in Fig.~\ref{softomega} where $\Omega(k)$ goes to a finite value as
$k\to 0$.  There is also no soliton feature at intermediate $k^*$.
For $N_b=32$ the dispersion $\Omega(k)$ has minima only at $k=0,\pi$.

\begin{figure}
\psfig{file=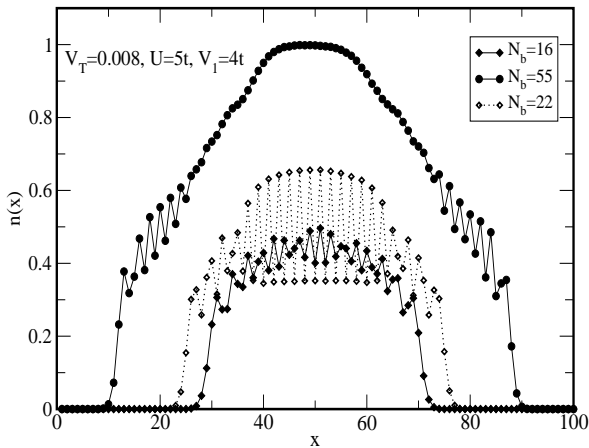,height=3in,width=3.5in,angle=-90}
\vskip-0.3in
\caption{Density profiles in a trap: For all fillings we see CDW
  oscillations. For $N_b=22$ the central region has long range CDW
  order while $N_b=55$ the central region is a MI. The $N_b=16$ case
  exhibits solitonic oscillations.}
\label{softtrapprof}
\end{figure}

\begin{figure}
\psfig{file=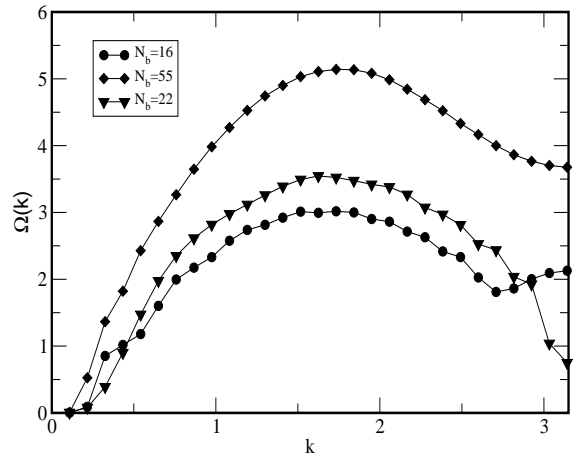,height=3in,width=3.5in,angle=-90}
\vskip-0.3in
\caption{The excitation energy, $\Omega(k)$, vs $k$ for the systems
  shown in Fig.~\ref{softtrapprof}. For $N_b=16$ there is a soliton
  excitation at $k<\pi$ as in the doped uniform system while for
  $N_b=22$, $\Omega(k^*=\pi)\to 0$ indicating the establishment of
  long range CDW order.}
\label{softtrapomega}
\end{figure}

\begin{figure}
\psfig{file=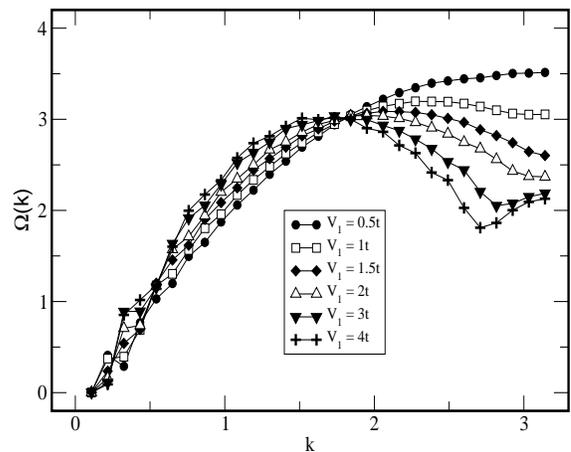,height=3in,width=3.5in,angle=-90}
\vskip-0.3in
\caption{Evolution of the excitation energy, $\Omega(k)$, with
  increasing $V_1$ for $U=5t$, $N_b=16$, $V_T=0.008$. The presence of
  solitonic excitations is clear for $V_1=3t$ and $4t$.}
\label{omegaNb16}
\end{figure}

\begin{figure}
\psfig{file=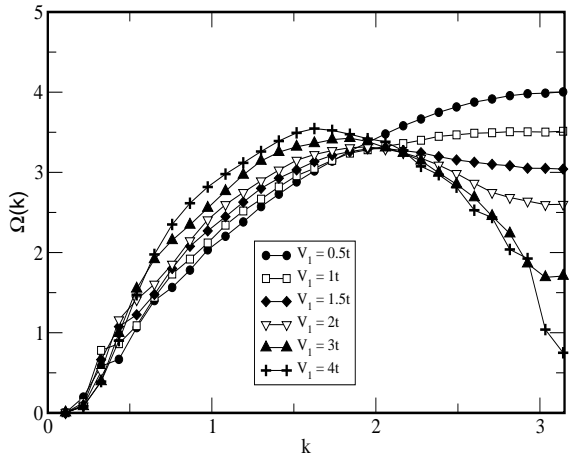,height=3in,width=3.5in,angle=-90}
\vskip-0.3in
\caption{Evolution of the excitation energy, $\Omega(k)$, with
  increasing $V_1$ for $U=5$, $N_b=22$, $V_T=0.008$. The gradual
  establishment of long range CDW order is clear.}
\label{omegaNb22}
\end{figure}

Placing the system in a trap destroys translational invariance.
Nonetheless, we shall now show that robust CDW and solitonic
excitations are observed. In Fig.~\ref{softtrapprof} the local density
profiles in a trap, $V_T=0.008$, are given for three fillings. For
$N_b=16$, solitonic oscillations are again evident (see also
Fig.~\ref{softtrapomega}) as local CDW correlations modulated by a
beat envelope.  For $N_b=22$ long range CDW order dominates although
some residual solitonic excitations remain near the edges. For
$N_b=55$ one sees a remarkable co-existence of several phases: CDW
towards the edges, followed by superfluid (no CDW oscillation and
compressible) and then a central incompressible Mott insulator (MI).
The density fluctuations in the two CDW regions are decoupled by the
intervening MI.  Such striking density oscillations have been observed
in non-neutral plasmas~\cite{plasma1,plasma2} which, due to their
electric charge, have long range repulsive interactions.

Figure~\ref{softtrapomega} shows $\Omega(k)$ for the same
fillings as in Fig.~\ref{softtrapprof}. For $N_b=16$ there
is a clear solitonic excitation of the type seen for the uniform
system, $k^*<\pi$. For the higher filling, $N_b=55$, the excitation
has shifted towards $k^*=\pi$ but $\Omega(k)$ remains relatively high,
indicating that this is not true long range order (as is of course
clear from the density profile). For $N_b=22$, on the other hand, we
see that $\Omega(k^*=\pi)$ is close to zero indicating the
establishment of long range CDW order.
Furthermore, there is no
evidence of a gap in $\Omega(k)$ as is seen in the uniform system at
half filling (see Fig.~\ref{softomega}): The system as a whole is
always compressible~\cite{batrouni02,wessel04}.

Finally, the evolution of the dispersion relation with increasing
near-neighbor repulsion is shown for $N_b=16$ in Fig.~\ref{omegaNb16}
and $N_b=22$ in Fig.~\ref{omegaNb22}.  In the former case, a soliton
minimum develops, while in the latter case CDW formation takes place
instead.

A further interesting feature of Fig.~\ref{omegaNb16}, with its
solitonic excitations, is the universal crossing of the dispersion
curves for different interaction strengths.  On the other hand, the
crossing in Fig.~\ref{omegaNb22}, where CDW order dominates, is not
universal. A similar well defined crossing point in the specific heat
has been seen both experimentally in $^3$He~\cite{brewer59} and in
fermion Hubbard models.\cite{georges93,vollhardt97,paiva01} We believe
a similar reasoning for the existence of crossing applies here.  The
integral of the structure factor over all momentum is constrained by
the density.  Thus if $S(k)$ increases with $V_1$ for some momenta
(for example at $k=\pi$ as CDW correlations build up), there must be a
corresponding decrease in $S(k)$ for other momenta.  This implies a
similar behavior in the dispersion relation $\Omega(k)$ and hence
suggests that dispersion curves for different interaction strengths
should cross.  As discussed in~\cite{vollhardt97} the universality of
the crossing in the specific heat case is a second, and more subtle
issue.

In conclusion, we have demonstrated that soliton signatures, which are
to be expected in the $d=1$ hard-core boson system owing to its close
connection with lattice fermion and spin models, are still very robust
when the bosons become soft-core and when they are placed in a
confining potential.  Rapid progress in the creation of near neighbor
repulsion $V_1$ in optical trap systems suggests that it will soon be
possible to look for these solitons experimentally.

We have also found that trapped bosons with near neighbor repulsion
can exhibit a remarkably rich density profile in which a Mott
insulator at commensurate filling occupied the trap center, followed
by a superfluid region and then a CDW region where the density is
locally pinned at $\frac12$, with a second, and final superfluid
region at the end of the occupied sites.  The local
compressibility~\cite{batrouni02,wessel04} also exhibits some unusual
features.  We find~\cite{batrouni05} that the CDW region is the most
compressible followed by the SF phase, in sharp contrast to a uniform
CDW which has a gap to charge excitations set by the near-neighbor
repulsion $V_1$. Experiments can measure $S(k)$ and therefore
$\Omega(k)$~\cite{steinhauer} which would serve to verify the presence
of solitons or other kinds of order. Similarly, as commented earlier,
the sound velocity is given by the linear slope at small $k$.


\noindent 
\underbar{Acknowledgements:}
We thank R. Kaiser and T. D. Night for very interesting and useful
discussions.  R.T.S. acknowledges support from NSF DMR 0312261, NSF
INT 0124863, and NSF ITR 0313390.


\end{document}